\documentclass[sigconf,nonacm]{acmart} 

\setcopyright{acmcopyright}
\copyrightyear{2018}
\acmYear{2018}
\acmDOI{10.1145/1122445.1122456}

\acmConference[Woodstock '18]{Woodstock '18: ACM Symposium on Neural
  Gaze Detection}{June 03--05, 2018}{Woodstock, NY}
\acmBooktitle{Woodstock '18: ACM Symposium on Neural Gaze Detection,
  June 03--05, 2018, Woodstock, NY}
\acmPrice{15.00}
\acmISBN{978-1-4503-XXXX-X/18/06}

\definecolor{bleu}{RGB}{68,114,196}
\definecolor{vert}{RGB}{0,128,0}

\begin{document}

\title{The DELICES project: Indexing scientific literature through semantic expansion}


\author{Florian Boudin}
\affiliation{%
 \institution{LS2N, Université de Nantes}
 \city{Nantes}
 \country{France}}
\email{florian.boudin@univ-nantes.fr}
 
\author{Béatrice Daille}
\affiliation{%
  \institution{LS2N, Université de Nantes}
  \city{Nantes}
  \country{France}
}
\email{beatrice.daille@univ-nantes.fr}

\author{Evelyne Jacquey}
\affiliation{%
  \institution{CNRS, Université de Lorraine}
  \city{Nancy}
  \country{France}}
\email{evelyne.jacquey@atilf.fr}

\author{Jian-Yun Nie}
\affiliation{%
  \institution{RALI, Université de Montréal}
  \city{Montréal}
  \country{Canada}}
\email{nie@iro.umontreal.ca}

\renewcommand{\shortauthors}{Boudin, et al.}

\begin{abstract}
Scientific digital libraries play a critical role in the development and dissemination of scientific literature.
Despite dedicated search engines, retrieving relevant publications from the ever-growing body of scientific literature remains challenging and time-consuming.
Indexing scientific articles is indeed a difficult matter, and current models solely rely on a small portion of the articles (title and abstract) and on author-assigned keyphrases when available.
This results in a frustratingly limited access to scientific knowledge.
%
%
The goal of the DELICES project is to address this pitfall by exploiting semantic relations between scientific articles to both improve and enrich indexing.
To this end, we will rely on the latest advances in semantic representations to both increase the relevance of keyphrases extracted from the documents, and extend indexing to new terms borrowed from semantically similar documents.
\end{abstract}

\ccsdesc[500]{Information systems~Digital libraries and archives}
\ccsdesc[500]{Information systems~Information retrieval}


\maketitle

\section{Introduction}

Scientific digital libraries (e.g.~arXiv, ACM Digital Library) play a critical role in the development and dissemination of scientific literature.
They provide researchers with access to millions of scientific articles, as well as an effective way to communicate their findings.
The picture, however, is not so rosy when it comes to searching and navigating through this ever-growing body of scientific literature.
Indeed, even with dedicated search engines, retrieving relevant publications is becoming increasingly challenging and time-consuming.
There are two main reasons for this situation.
First and foremost, the global explosion of the amount of scientific output is overwhelming researchers with an unsustainable torrent of publications. 
The exponential growth of the number of submissions in arXiv is a very compelling illustration of that issue.\footnote{\url{https://arxiv.org/stats/monthly_submissions}}

\let\thefootnote\relax\footnotetext{"Copyright © 2020 for this paper by its authors. Use permitted under Creative Commons License Attribution 4.0 International (CC BY 4.0)."}

The second reason is that the current practice in indexing scientific articles typically relies on a small fraction of their content (title and abstract), which makes it ineffective~\cite{7809016}.
This is not only due to licensing issues, in the case of commercial publishers, but also to resource limitations in the case of public and preprint repositories~\cite{10.1145/3295750.3298953}.
%
One straightforward solution to alleviate this problem is to supplement paper indexing with keyphrases, that is, single or multi-word lexical units that capture the main topics of a document~\cite{barker1972comparative,zhai-1997-fast,Gutwin:1999:IBD:338985.338996}.
Most documents, however, do not come with associated keyphrases, and manual annotation is simply not a feasible option~\cite{Mao2017}.
Needless to say, there is an acute need for automated keyphrase assignment models, task on which researchers both in information retrieval and natural language processing are now devoting their efforts.

\begin{figure}[htb!]
    \centering
    \includegraphics[width=.39\textwidth]{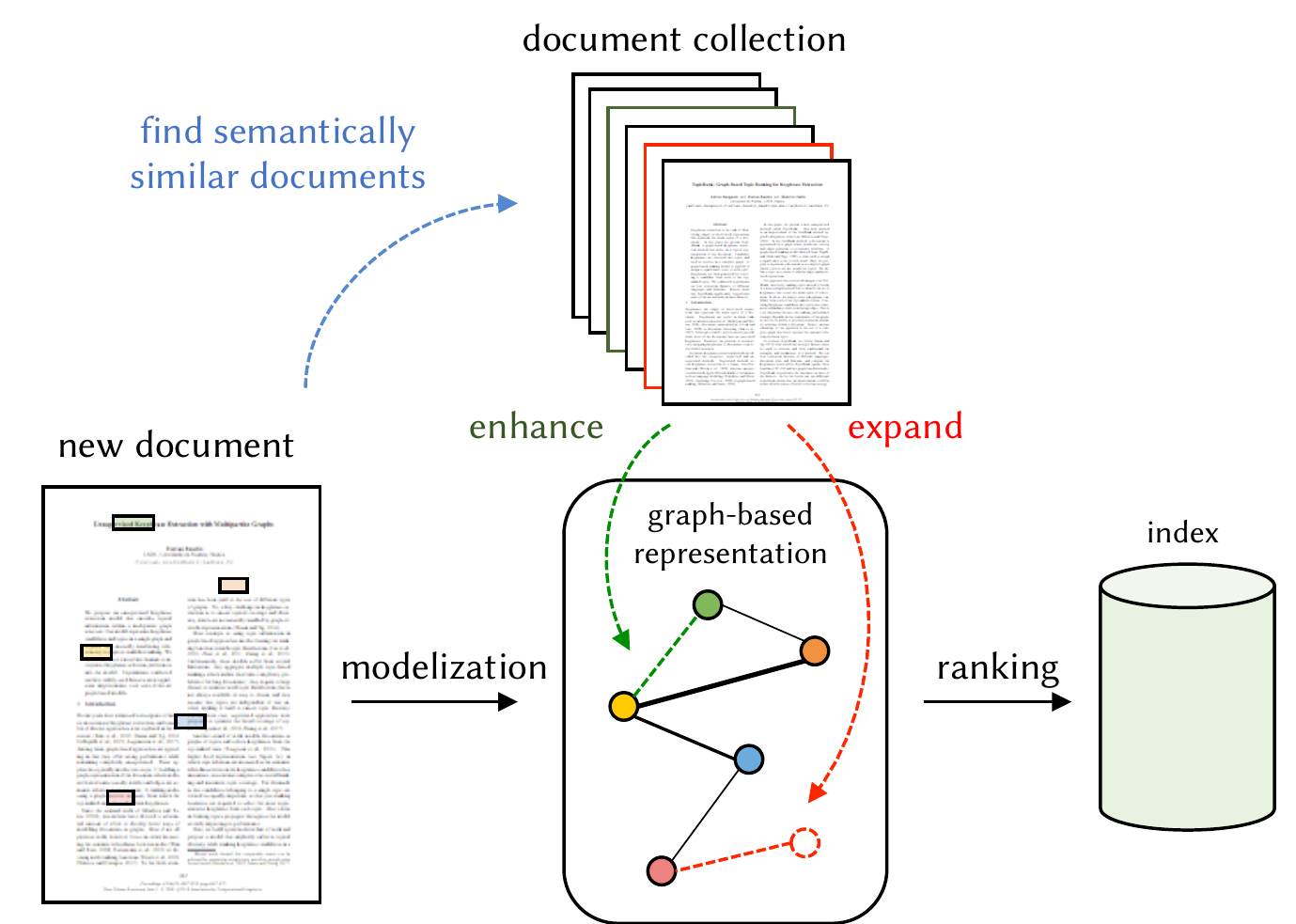}
    \caption{Overall architecture of the indexing model.}
    \label{fig:architecture}
\end{figure}

Despite the higher performance brought by the use of neural network architectures, current models for identifying keyphrases still achieve fairly low precision scores~\cite{P17-1054,zhao-zhang-2019-incorporating}.
The primary reason behind this is their inability to accurately produce keyphrases that do not occur in the documents.
%
%
Those so-called \textit{absent} keyphrases are especially valuable for indexing documents, accounting for about half of the manually assigned keyphrases~\cite{hasan-ng:2014:P14-1}.
Overlooking these absent keyphrases ultimately results in relevant documents that are not retrieved, thus preventing a thorough exploration of the scientific knowledge.
%
The goal of the DELICES project is to provide a solution to this critical problem by exploiting the relationships between scientific articles to improve and enrich indexing.
More precisely, as depicted in Figure~\ref{fig:architecture}, we will rely on \textcolor{bleu}{automatically detected, semantically similar documents} to both \textcolor{vert}{enhance the weighting scheme} for keyphrases that occur in the documents and \textcolor{red}{extend the index with new terms} borrowed from similar documents.

The unique nature of the DELICES project comes from the underlying graph-based ranking model that sits at its core, which allows for the flexible integration of prior domain knowledge that is required to accurately predict absent keyphrases.
This is motivated by the fact that current neural-based models not only have difficulties harnessing such knowledge, but also exhibit poor generalization performance across domains which severely limits their use in practice~\cite{gallina-etal-2019-kptimes}. 
Another important aspect of this project is the automatic acquisition of the prior domain knowledge through the compilation of sets of semantically related documents.
These latter will enable us to enrich the graph representations of the documents, facilitating the overall ranking process~\cite{Wan:2008:SDK:1620163.1620205} while allowing the prediction of absent keyphrases that only occur in related documents. 
The rest of this paper discusses the two main challenges we aim to address in DELICES, and concludes with a summary of the work already accomplished since the project started.
 
\section{Challenges}

\subsection{Improving keyphrase ranking}

The content of scientific papers alone is often not sufficient to effectively rank keyphrases~\cite{Wan:2008:SDK:1620163.1620205,hasan-ng:2014:P14-1} -- an issue that is exacerbated by the limited availability of full-text articles.
It is thus essential that external knowledge can be made accessible to the model so that it can make better predictions.
That being said, selecting the appropriate amount of knowledge and integrating it into the keyphrase generation model is not straightforward, as this process can easily cause the indexation to drift away from the original content of the documents if not done carefully.
To avoid this pitfall, we will seek to identify precisely where, in the graph representation, more information is needed, and devise a fine-grained enrichment mechanism that can efficiently and reliably improve the overall ranking of keyphrases.
A first step in this direction would be to act on the weakly connected components of the graph by introducing and strengthening edges between keyphrase candidates that co-occur within the related documents.

\subsection{Generating absent keyphrases}

As stated earlier, absent keyphrases are of utmost importance when indexing scientific articles~\cite{hasan-ng:2014:P14-1,P17-1054} -- they act as a means for expanding documents, and thus alleviate the ``vocabulary mismatch'' problem between query terms and relevant documents.
Predicting appropriate absent keyphrases is obviously a very challenging task because of the unrestricted search space: every possible term can be considered as keyphrase.
It therefore comes as no surprise that recent neural keyphrase generation models still achieve fairly low accuracies on that task despite being trained on large amounts of annotated data~\cite{P17-1054,zhao-zhang-2019-incorporating}.
Furthermore, their predictions are bound to a fixed-size output vocabulary built from the set of gold standard keyphrases, which means that they can only produce keyphrases that were already assigned to other documents.
With this in mind, we are looking to move away from these data hungry models, and study how absent keyphrases could be inferred from related documents present in the entire collection.
Here we argue that carefully selected sets of semantically related documents are the right spot for mining new, yet likely to be relevant indexing terms.
To verify this claim, a straightforward approach would be to expand the document representation with these new terms in order for the model to predict absent keyphrases.
The main difficulty with this approach would be to precisely control how present and absent keyphrases are interleaved in the overall ranking.
We believe that jointly encoding the document and domain knowledge into a multigraph data structure would allow for a finer-grained weighting of the keyphrases.

\section{Preliminary results}

One critical issue with the current literature on keyphrase assignment is the lack of unified evaluation methodology, making the direct comparison between previous models not possible.
In an attempt to solve that issue, we conducted the first large-scale analysis of state-of-the-art keyphrase extraction models involving multiple benchmark datasets from various sources and domains~\cite{gallina2020large}.
Our main results reveal that existing models are still challenged by simple baselines on some datasets, and yield insights about the negative impact of using author-assigned keyphrases as a proxy for gold standard.
We also provide specific recommendations on which baselines and datasets should be included in future work.
%
%
%


Neural models for keyphrase generation do not generalize well across domain, but the extent to which their performances are degraded is not clearly understood as only one sufficiently large training dataset was available~\cite{P17-1054}.
To fill that gap, we collected a new large-scale dataset, namely KPTimes, composed of news texts paired with editor-curated keyphrases.
Using it, we provided an in-depth analysis of the performance of state-of-the-art neural models and investigated their transferability to the news domain as well as the impact of domain shift~\cite{gallina-etal-2019-kptimes}.

%
%

\begin{acks}
We thank the anonymous reviewers for their valuable comments.
This work was supported by the French National Research Agency (ANR) through the DELICES project (ANR-19-CE38-0005-01).
\end{acks}

\bibliographystyle{ACM-Reference-Format}
\bibliography{references}

\end{document}